# SAFETY ENHANCEMENT THROUGH SITUATION-AWARE USER INTERFACES

*V. De Florio, C. Blondia*

*PATS research group, Universiteit Antwerpen & IBBT*
*Middelheimlaan 1, 2020 Antwerp, Belgium*
vincenzo.deflorio@ua.ac.be

## Abstract

Due to their privileged position halfway the physical and the cyber universe, user interfaces may play an important role in learning, preventing, and tolerating scenarios affecting the safety of the mission and the user's quality of experience. This vision is embodied here in the main ideas and a proof-of-concepts implementation of user interfaces combining dynamic profiling with context- and situation-awareness and autonomic software adaptation.

## 1 Introduction

The user interface (UI) may be considered as the contact point between two "universes"—the physical universe of the user (let us refer to this universe as **U**) and the cyber universe where required computer services are executed (**C**). The UI is also the logical "place" where actions are selected and passed for execution in **C**. As well known **U** and **C** are very different from each other—in particular they have quite different notions of time, behaviours, actions, and quality of service. Despite so huge a difference, the consequences of the actions in **C** often reverberate in **U**—to the point that when the computer service is safety-critical failures or misinterpretations in **C** may induce catastrophic events in **U** possibly involving the loss of goods, capital, and even lives. As a matter of facts, the human factor is known as one of the major causes for system failures [2,15], and the UI is often the indirect player behind most interaction faults at the root of computer failures.

Due to its central role in the emergence of the user's quality of experience (QoE), the UI has been the subject of extensive research. As a result, current interfaces are adaptive, anticipative, personalized, and to some degree "intelligent" [3].

We believe that much more can be done beyond this already noteworthy progress. Thanks to its privileged position halfway between the user and the computer, we argue that the UI is well suited for hosting several non-functional tasks, including:

- Gathering contextual information from both sides of the activity spectrum.
- Deriving situational information about the current interaction processes.
- Producing logs of the knowledge accrued and situations unveiled.
- Executing corrective actions in **U** and **C** so as to mitigate the extent of the consequences of safety or security violations.

In this paper we propose an approach based on the above argument. This approach instruments a UI so as to produce a stream of atomic UI operations and their **C**-time of occurrence—such as the user typing a key at clock tick $t_1$ or selecting a widget at clock tick $t_2$. This stream is transcoded into a suitable form (we call it an "interaction code", or iCode for short) that can be logged for post-mortem analysis and/or used for run-time analysis, e.g. dynamically building or refining a model of **U**. Well-known techniques such as stereotypes, rules, hidden Markov models, or Bayesian intelligence [3] may then be used e.g. to characterize the expected, "normal" behaviour of a group of authorized "super users" of a safety critical system. When such behaviours would be represented in an adequate form, the UI could then be instructed to function as a sort of custom biometric sensor and recognize whether the current usability patterns correspond to one of these users. For instance, when a stereotype would change dynamically from that of a known user to a different one, this may signal the occurrence of several new situations, including:

1. The user is no more in command (e.g. due to stress, fatigue, drugs, or other external situations).
2. The user has changed (e.g. because a new person took control after the authorized person logged in, or because of a cyber attack).
3. The likely occurrence of a performance failures, i.e. a violation of an agreed-upon scheduling of events involving both **C** and **U**.

Correspondingly, UI may tolerate situations such as 1 and 2 by raising local/global alerts and/or shutting down critical functionality in the UI. The corresponding loss in availability would be the price to pay for the enhancement in safety. Detecting a loss of responsiveness jeopardising the real-time specifications of the service could be avoided by defining safe default actions to be taken autonomously when the probability of performance failures arise.

In this paper we present a prototypic demonstrator meant as a proof of feasibility for UI-based biometric sensors. Our system, embedded in a very simple Tcl/Tk [14] UI, is currently based on naïve rules able to detect a few QoE failures. When such simple context changes are detected, we show how our UI can take simple forms of autonomic adaptation during the system run-time. These adaptations may be used e.g. to adjust dynamically the privileges of the current user or request a new verification of the user's identity, which paves the way to full-fledged safety enforcement protocols.

This paper is structured as follows: in Sect. 2 we present the main assumptions and the basic elements of our approach. Section 3 deals with context collection and analysis – the foundation layer of our approach. Context and situation analysis and planning are the focus of Sect. 4. Section 5 briefly describes how adaptations are enacted. Our conclusions are finally summarized in Sect. 6.

## 2  Assumptions and Approach

Our approach is based on the classic steps of autonomic computing [9], specialised and *adapted* as follows:

1. *Collection* of the user-context, *viz.* all those **U** actions that result from expected and unexpected human-computer interaction, as well as their times of occurrence. Context collection may be interpreted as the perception layer of our approach [5].
2. *Context analysis*, that is analysis of the collected actions and times meant to derive simple knowledge about the behaviour and the state of the current user, and in particular some indication of the current QoE.
3. *Situation analysis and identification*, that we interpret here simply[1] as the analysis of multiple instances of the context knowledge derived in step 2 so as to detect the onset of complex high-level situations [18] requiring proper corrective responses. The term apperception is used to refer to both context and situation analysis [5].
4. *Planning* of corrective responses to the onset of situation or the detection of QoE losses. In practice, this step requires the definition of software evolutions matching the current user-context and meant to optimize QoE and tolerate situations threatening the mission design goals and the validity of the system assumptions [6]. Evolution engine is the term used in [4] to refer to a system's ability to plan corrective responses.
5. *Execution* of the selected software evolution.

In what follows we introduce the solutions we adopted to implement those five steps.

---

1  For a more general discussion of situation analysis we refer the reader to [18].

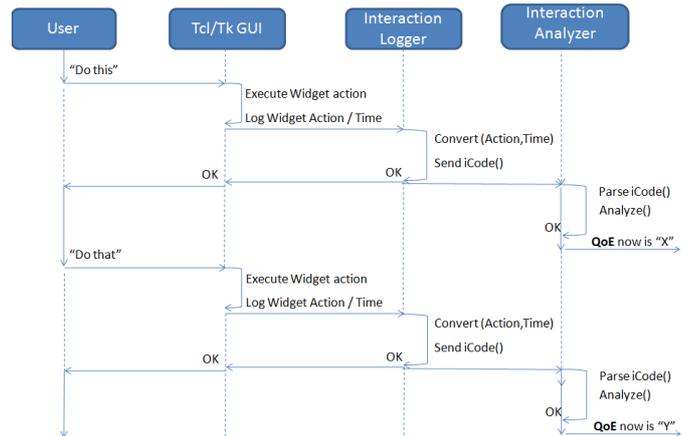

Figure 1. Sequence diagram of *Collection* and *Analysis* steps.

## 3  Context Collection and Analysis

The *Context collection and analysis* steps in our strategy are better described via Fig. 1. Four main "actors" are portrayed therein: the user in universe **U** produces events by interacting with a Tcl/Tk [14] graphical UI (GUI). In addition to its normal processing our GUI executes an extra task transparently of the user: for each action being executed or activity being sensed, the GUI requests an Interaction Logger to produce a string describing that action / activity and a time-stamp reporting the amount of milliseconds elapsed since a computer-chosen epoch. We refer to such strings as to "the iCode". The iCode may be considered as a concise representation of GUI events reflecting some of the use-time events. The choice of which events to reflect is an important design choice depending on technological, design, mission, and cost constraints and influencing the quality of the ensuing inferences, which are carried out by a component called Interaction Analyzer (IA). Aim of the Interaction Analyzer is parsing the iCode—possibly producing a context knowledge base—and analysing the interaction e.g. looking for misbehaviours or other signs of non-optimal QoE. This analysis could for instance infer that the user is in full command of the operations, or contrariwise that he or she is experiencing difficulties in operating the system. Reactions to this could range from distress signals to fail-safe enforcement, or just be used by the GUI itself to adjust autonomically its look-and-feel and features.

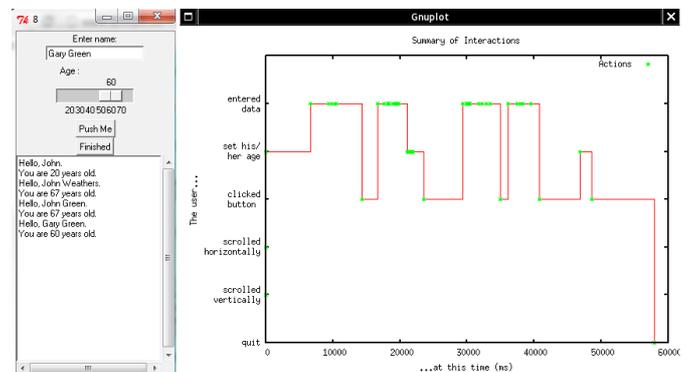

Figure 2. Summary of the interactions carried out through a simple Tcl/Tk GUI.

A practical example of our *Collection* & *Context Analysis* strategy is shown in Fig. 2. The picture shows on the left hand side a snapshot from the interactions between a user and a Tcl/Tk GUI. The right hand side picture is a summary of all interactions from the moment the GUI is invoked until the user presses the "Finished" button. As can be seen, the ordinates represent a set of GUI actions while the abscissas state the times of occurrence of those actions. In this particular example, the user first sets his age, then enters data, then clicks the "Push Me" button; after this, he edits data, sets age again, then clicks "Push Me"; and so forth. Timing is also quite as it could be expected by an average user, with sudden editing actions followed by periods of inactivity.

Figure 3. Excerpts from the source code and iCode of the GUI shown in Fig. 2.

Figure 3 shows excerpts from the GUI code and the iCode corresponding to the interactions of Fig. 2.

As anticipated in Fig. 1, a lexical analyser parses the iCode and executes one or more analyses (see Fig. 4). The nature and the extent of those analyses constitute a research topic in itself and are likely to link together as different a discipline as artificial intelligence, Bayesian intelligence, situation calculus, and temporal logic—to name just a few [3]. The ensuing results would then be reflected in several "biometric sensors" as described e.g. in [7,8]. Our preliminary results are based on two illustrative analyses shown in next section.

Figure 4. Excerpts from the iCode parser.

Analyses are carried out by functions that are called, one by one, at the end of the parsing phase. As mentioned already, various strategies may be employed, though the illustrative ones we used here are based on the inspection of two stacks of integer variables, called actstack and timestack and first shown in Fig. 4. These stacks are filled in during the lexical analysis in order to serialize iCode actions and their corresponding times of occurrence. By inspecting these stacks it is possible to detect several cases of discomfort or misuse. In what follows we describe three examples by means of excerpts from their source code (Fig. 5).

The first function in Fig. 5 checks whether the "Push Me" button had been pressed before any input entry had occurred. The second function checks whether multiple consecutive occurrences of the same action ("Push Me") are present in the stack. The third one measures the time gradient while typing in the input field. A typing rate of more than three characters per second is assumed to be senseless and associated with some source of distress. More complex analyses are likely to require more sophisticated temporal-related techniques—again, a matter in itself worth of research attention.

Figure 5. Discomfort detection: Three examples.

Figure 6 and Fig. 7 provide a summary of the interactions during two runs of our Tcl/Tk interface, including three discomfort detections.

As mentioned before, discomfort detections may trigger corrective measures—for instance, a reshaping of the UI or a request to reassert the user's identity. Such corrective measures should take other context properties into account, which could be represented as context models: sets of assumptions characterizing the context the UI is meant to be deployed in [6]. Context models could be as simple as a device model and consist of e.g. a screen's dimensions, or they could include complex assertions about the target user—e.g. whether the user is impaired or not, or whether he or she is accustomed to computer technology or otherwise. The design of domain-specific context models should be conducted with domain experts—e.g., in the case of target user context models, by psychologists or sociologists able to capture and express the essence of what the target user would consider as positive QoE [3,16].

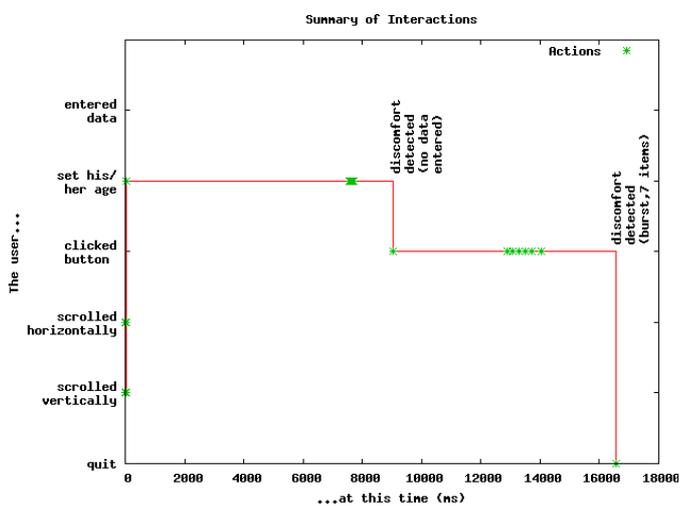

Figure 6. Summary of interactions, with two discomfort detections.

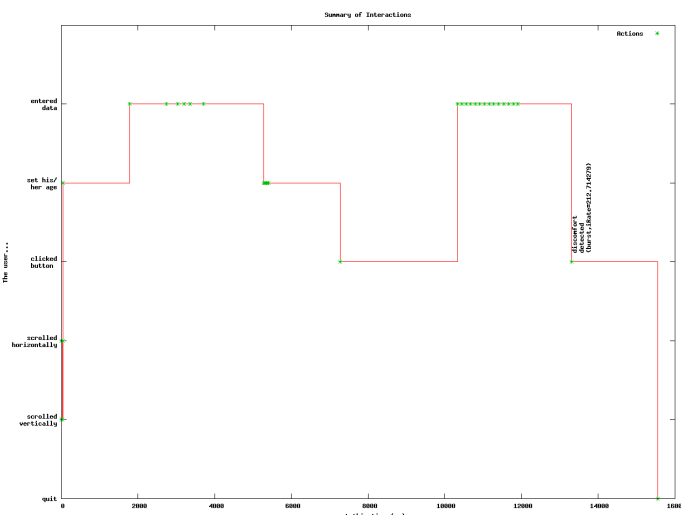

Figure 7. Summary of interactions, with a third type of discomfort detection. In this case during a burst of input actions the user typed in an average of 4.7 characters per second. As this is well above our threshold (3 chars/sec), a case of suspected discomfort is declared.

## 4 Analyses and Planning

When the GUI mission is confined to optimizing the current user's QoE, *Context analysis* may be sufficient to direct the ensuing *Planning* step. An example of this is described in Fig. 8, in which a simple GUI adaptation is carried out to adjust optimally the size of a widget. In this case we made use of a simple technological context model—a screen size model: the screen size is assumed to be that of a hand-held device with a resolution of 260×$H$ pixels and $H$>260. The case is described in Fig. 8 and results in a progressive adjustment of a Scale widget [14]. Once the widget becomes too large to fit in horizontal mode, the UI is automatically adapted and the widget is set in vertical mode.

In some other cases *Situation analysis and identification* (SAI) as a higher level form of analysis may prove to be indispensable in order to take higher level decisions—such as those pertaining to system safety and security. In fact SAI allows **C** or UI events and states to be put in relation with complex **U** situations. A thorough discussion on this situation identification and related techniques may be found in [18]. A very simple example of this is the use of rules stating the onset of **U** situations after the occurrence of several UI events —for instance, situation $s_1$ = "User is likely to have changed" may be declared after a given number of consecutive erroneous or senseless sequences of UI operations. Another example would be situation $s_2$ = "User is likely to have been taken over by a computer". Once situations such as $s_1$ or $s_2$ are declared, the *Planning* and *Execution* steps enforce some form of corrective adaptation of the UI meant to guarantee the integrity of the mission. An example of this is shown in Fig. 9. In this case the onset of situation $s_1$ triggers a temporary disabling of functionality of the UI until the user has re-entered his or her credentials. Other forms of protection may involve raising alarms when suspicion periods are started and producing "black-box"-like logs of activities for off-line analysis of the performance of the operators. Protection against situations such as $s_2$ may involve different strategies e.g. a request to solve a CAPTCHA [1].

## 5 The Execution Step

When embedded in a GUI, execution requires an evolution of the structure and functions of the user interface. In our prototypic implementation this is carried out by embedding the analysis in the GUI itself, which of course means intertwining the functional and the adaptation concerns.
Another approach was carried out in [17], in which a Java GUI is serialized, evolved, and then reloaded to execute the planned adaptations. This allowed a straightforward implementation of what we refer to as *widget paging*—a technique that manages the widgets in a GUI similarly to memory pages in virtual memory. In so doing the screen space is allocated to widgets in a way that reflects the frequency of usage of widgets, with the least frequently used widgets removed from the screen. Figure 11 shows an experiment with a winner-takes-all strategy—the most frequently used widget taking all the screen space.

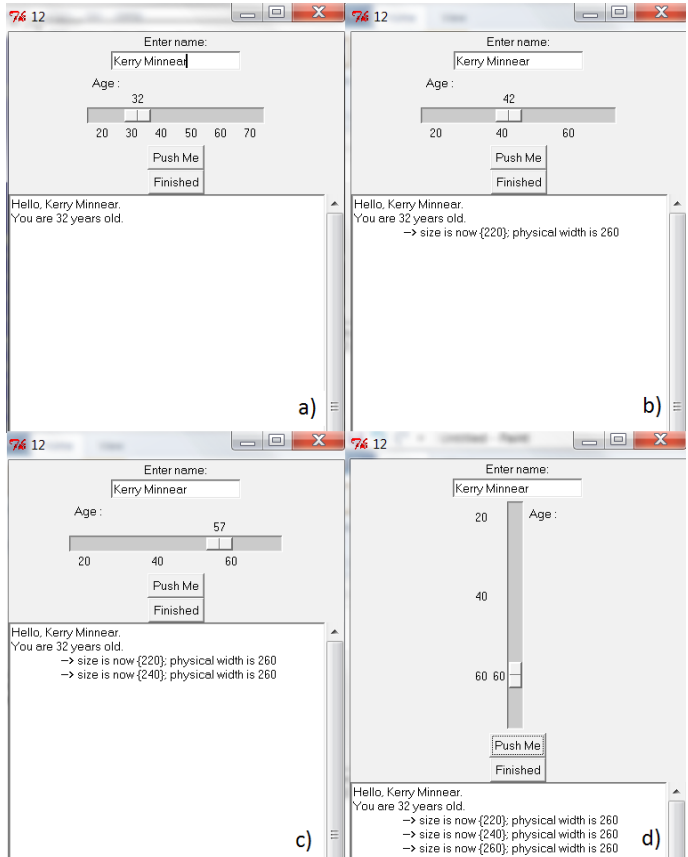

usually is the inert witness of behaviours and situations that, when correctly understood, could trigger actions to mitigate the extent and severity of those failures—if not tolerate them. By making user interfaces aware of those dynamic situations and context changes we argued that it may be possible to enhance at the same time system safety, usability, and quality of experience. In this paper we have introduced the main concepts and a prototypic implementation of user interfaces compliant to this vision and coupling dynamic profiling [3], situation awareness [18], and autonomic computing [9].

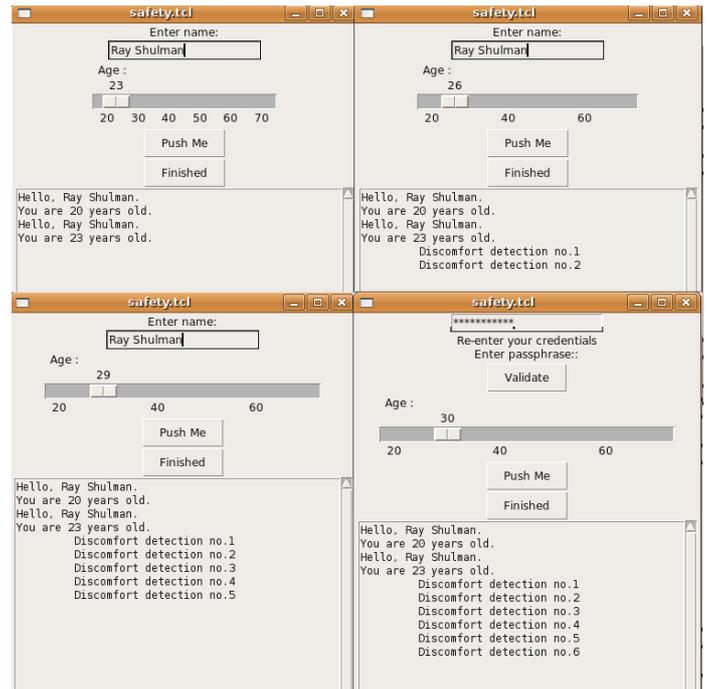

Figure 8. Adaptation actions triggered by rapid bursts of "scale" actions. The GUI is deployed in a screen with a width of 260 units (on a conventional PC screen, about 6cm). Height is assumed to be greater than 260 units. The Scale widget is initially set to a width of 200 units. In a) no adaptation is carried out. In b) and c) as a result of two bursts of "scale" actions the size of the widget reaches 240 units. A further burst in d) triggers a change in orientation of the widget, which now is set to a *height* of 260 units.

Extensively used in software adaptation, aspect orientation [13] is a well-known technique that can be used to achieve effectively separation of design concerns in the *Execution* step of our approach. Aspects allow adaptation logics to be modularized as individually deployable units that can be directly weaved in the application business logics either off-line or during the run-time [12], which makes aspect orientation an ideal tool to realize systems such as the one reported in this paper. One major drawback is that it calls for specific linguistic support, which is not available for Tcl/Tk. An interesting alternative is given by Transformer, a Java / OSGi framework for adaptation behaviour composition that dynamically selects and merges reusable and adaptation modules in function of the current context [10,11,12]. Finally, the *Execution* step may be realized as callbacks on so-called reflective and refractive variables [7,8].

## 6 Conclusions

Human errors are at the core of many a catastrophic failure. Such errors often take place behind a user interface, which

```
proc react {} {
   global sz
   global phy_w

   # add "Exit" to a temporary copy of "foo.txt"
   exec cp foo.txt footmp.txt
   set chan [open footmp.txt a]
   puts $chan "button(Exit) @[clock clicks -milliseconds]"
   close $chan

   set rv [ catch { exec anagui < footmp.txt } msg ]
   # .txt insert end "\t--> Script returned {$msg}\n"
   set msg [ string trimleft $msg ]
   set msg [ string range $msg 0 0 ]
   # .txt insert end "\t--> trimmed return value {$msg}\n"

   if { $msg == "8" } {
   . . . . .
```

Figure 10. A trivial way to enact the Execution adaptation step. Procedure react is called each time a situation or context change is detected. In this case there is no separation of the design concerns.

Despite the simplistic design of our proof of concepts, our system already allows several behaviours to be tracked turning a user interface into a simple and cheap usability and biometric sensor. When coupled with a framework such as Transformer [10,11,12] our system may also allow the computer system to be reshaped after the dynamic model of the current user, preventing unnecessary or unsafe functionality to be "unloaded" from both the system and the user interface. This may also be used to produce "big data" about the usability of computer services and tell enterprises what features are the most desired or the most hated in their product. Yet another by-product could be the realisation of design-for-all interfaces based on a WYSIWYE ("what you see is what you expect") principle.

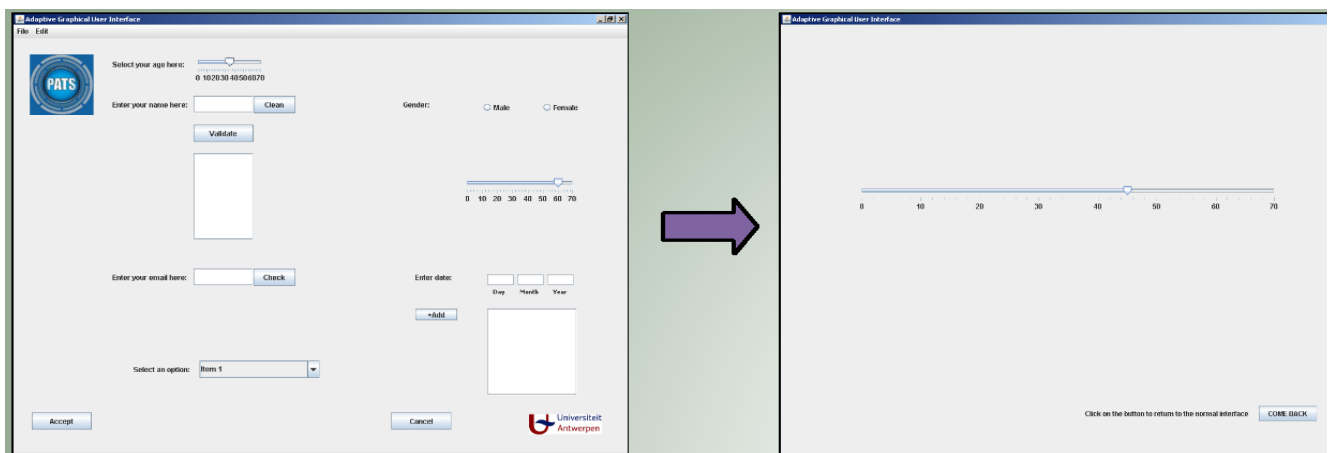

Figure 11. Widget paging with a single widget: the screen is temporarily allocated to the single widget the user focuses his or her attention to. At any time the user can restore the original structure by clicking a button.